\documentclass[english]{article}
\usepackage[latin9]{inputenc}
\usepackage{color}
\usepackage{amsbsy}
\usepackage{amssymb}

\newcommand{\lyxaddress}[1]{
\par {\raggedright #1
\vspace{1.4em}
\noindent\par}
}

\usepackage{babel}

\begin{document}

\title{\textbf{Gravitomagnetic effect in gravitational waves}}

\author{\textbf{$^{1}$Lorenzo Iorio and $^{2}$Christian Corda}}

\maketitle

\lyxaddress{\begin{center}
\textbf{$^{1}$}INFN - Sezione di Pisa, Largo Pontecorvo 3, I - 56127
PISA, Italy; \textbf{$^{2}$}Associazione Scientifica Galileo Galilei,
Via Pier Cironi 16 - 59100 PRATO, Italy
\par\end{center}}

\lyxaddress{\begin{center}
\textit{E-mail addresses:} \textbf{$^{1}$}\textcolor{blue}{lorenzo.iorio@libero.it;
}\textbf{$^{2}$}\textcolor{blue}{cordac.galilei@gmail.com}
\par\end{center}}
\begin{abstract}
After an introduction emphasizing the importance of the gravitomagnetic
effect in general relativity, with a resume of some space-based applications,
we discuss the so-called magnetic components of gravitational waves
(GWs), which have to be taken into account in the context of the total
response functions of interferometers for GWs propagating from arbitrary
directions. 
\end{abstract}

\lyxaddress{PACS numbers: 04.80.Nn, 04.80.-y, 04.25.Nx}

In the weak-field and slow motion approximation, the Einstein field
equations of general relativity, which establish  how the mass-energy distribution determines the spacetime metric, get linearized resembling to the Maxwellian
equations of electromagnetism. As a consequence, a ``gravitomagnetic''
field $\boldsymbol{B}_{g}$, induced by the off-diagonal components
$g_{0i},\mbox{ }i=1,2,3$ of the spacetime metric tensor related
to the mass-energy currents of the source of the gravitational field,
arises \cite{key-1}. The gravitomagnetic field affects orbiting test
particles, precessing gyroscopes, moving clocks and atoms and propagating
electromagnetic waves \cite{key-2,key-3}. Perhaps, the most famous
gravitomagnetic effects are the precession of the axis of a gyroscope
\cite{key-4,key-5} and the Lense-Thirring precessions of the orbit
of a test particle \cite{key-6}, both occurring in the field of a
central slowly rotating mass like, e.g., our planet. Direct, undisputable
measurements of such fundamental predictions of general relativity
are not yet available.

Some attempts to detect the Lense-Thirring effect have been more or
less recently performed in the gravitational fields of the Sun \cite{key-7},
Earth \cite{key-8,key-9,key-10} and Mars \cite{key-11} with natural
(the inner planets of the Solar System) and artificial bodies (the
terrestrial LAGEOS satellites and the martian Mars Global Surveyor
probe); some of them have raised debates concerning their reliability
and/or realistic level of accuracy reached \cite{key-12,key-13,key-14}.
The LARES satellite, recently approved by ASI with the claimed goal
of measuring the Lense-Thirring effect together with the exiting LAGEOS
and LAGEOS II at a $\approx1\%$ level, should be launched with a
VEGA rocket in 2010-2011, but doubts exist that it will effectively
be able to reach the expected level of accuracy \cite{key-15}.

The dedicated GP-B mission \cite{key-16,key-17}, aimed to measure
the Pugh-Schiff effect with four superconducting gyroscopes carried
on board a spacecraft in a polar orbit around the Earth has not (yet?)
obtained the expected accuracy ($1\%$ or better) because of the occurrence
of some unexpected competing systematic effects \cite{key-18,key-19,key-20}.

The gravitomagnetic field plays also a fundamental role  in some astrophysical scenarios like rotating black holes and neutron stars \cite{buco1,buco2,buco3}.

Recently, starting by the analysis in \cite{key-21}, some papers
in the literature have shown the importance of the gravitomagnetic
effects in the framework of the GWs detection too \cite{key-22,key-23,key-24}.
In fact, the so-called {}``magnetic'' components of GWs have to be
taken into account in the context of the total response functions
of interferometers for GWs propagating from arbitrary directions,
see \cite{key-24} for a review. In this proceeding paper, the interferometric
response functions for the magnetic components are re-analysed following
the lines of \cite{key-22}. As interferometric GWs detection is performed
in a laboratory environment on Earth, the coordinate system in which
the space-time is locally flat is typically used \cite{key-25} and
the distance between any two points is given simply by the difference
in their coordinates in the sense of Newtonian physics. In this frame,
called the frame of the local observer, GWs manifest themselves by
exerting tidal forces on the masses (the mirror and the beam-splitter
in the case of an interferometer). We work with $G=1$, $c=1$ and
$\hbar=1$ and we call $h_{+}(t_{tt}+z_{tt})$ and $h_{\times}(t_{tt}+z_{tt})$
the weak perturbations due to the $+$ and the $\times$ polarizations
which are expressed in terms of synchronous coordinates $t_{tt},x_{tt},y_{tt},z_{tt}$
in the transverse-traceless (TT) gauge. In this way, the most general
GW propagating in the $z_{tt}$ direction can be written in terms
of plane monochromatic waves \cite{key-22}

\begin{equation}
\begin{array}{c}
h_{\mu\nu}(t_{tt}+z_{tt})=h_{+}(t_{tt}+z_{tt})e_{\mu\nu}^{(+)}+h_{\times}(t_{tt}+z_{tt})e_{\mu\nu}^{(\times)}=\\
\\=h_{+0}\exp i\omega(t_{tt}+z_{tt})e_{\mu\nu}^{(+)}+h_{\times0}\exp i\omega(t_{tt}+z_{tt})e_{\mu\nu}^{(\times)},\end{array}\label{eq: onda generale}\end{equation}

and the correspondent line element will be

\begin{equation}
ds^{2}=dt_{tt}^{2}-dz_{tt}^{2}-(1+h_{+})dx_{tt}^{2}-(1-h_{+})dy_{tt}^{2}-2h_{\times}dx_{tt}dx_{tt}.\label{eq: metrica TT totale}\end{equation}

The wordlines $x_{tt},y_{tt},z_{tt}=const.$ are timelike geodesics
representing the histories of free test masses \cite{key-21,key-22}.
The coordinate transformation $x^{\alpha}=x^{\alpha}(x_{tt}^{\beta})$
from the TT coordinates to the frame of the local observer is \cite{key-22}

\begin{equation}
\begin{array}{c}
t=t_{tt}+\frac{1}{4}(x_{tt}^{2}-y_{tt}^{2})\dot{h}_{+}-\frac{1}{2}x_{tt}y_{tt}\dot{h}_{\times}\\
\\x=x_{tt}+\frac{1}{2}x_{tt}h_{+}-\frac{1}{2}y_{tt}h_{\times}+\frac{1}{2}x_{tt}z_{tt}\dot{h}_{+}-\frac{1}{2}y_{tt}z_{tt}\dot{h}_{\times}\\
\\y=y_{tt}+\frac{1}{2}y_{tt}h_{+}-\frac{1}{2}x_{tt}h_{\times}+\frac{1}{2}y_{tt}z_{tt}\dot{h}_{+}-\frac{1}{2}x_{tt}z_{tt}\dot{h}_{\times}\\
\\z=z_{tt}-\frac{1}{4}(x_{tt}^{2}-y_{tt}^{2})\dot{h}_{+}+\frac{1}{2}x_{tt}y_{tt}\dot{h}_{\times},\end{array}\label{eq: trasf. coord.}\end{equation}

where it is $\dot{h}_{+}\equiv\frac{\partial h_{+}}{\partial t}$
and $\dot{h}_{\times}\equiv\frac{\partial h_{\times}}{\partial t}$.
The coefficients of this transformation (components of the metric
and its first time derivative) are taken along the central wordline
of the local observer \cite{key-21,key-22}. It is well known from
\cite{key-21} that the linear and quadratic terms, as powers of $x_{tt}^{\alpha}$
, are unambiguously determined by the conditions of the frame of the
local observer, while the cubic and higher-order corrections are not
determined by these conditions. Thus, at high-frequencies, the expansion
in terms of higher-order corrections breaks down \cite{key-21,key-22}.
Considering a free mass riding on a timelike geodesic ($x=l_{1},$
$y=l_{2},$ $z=l_{3}$ ) \cite{key-21,key-22}, eqs. (\ref{eq: trasf. coord.})
define the motion of this mass with respect to the introduced frame
of the local observer. In concrete terms one gets

\begin{equation}
\begin{array}{c}
x(t)=l_{1}+\frac{1}{2}[l_{1}h_{+}(t)-l_{2}h_{\times}(t)]+\frac{1}{2}l_{1}l_{3}\dot{h}_{+}(t)+\frac{1}{2}l_{2}l_{3}\dot{h}_{\times}(t)\\
\\y(t)=l_{2}-\frac{1}{2}[l_{2}h_{+}(t)+l_{1}h_{\times}(t)]-\frac{1}{2}l_{2}l_{3}\dot{h}_{+}(t)+\frac{1}{2}l_{1}l_{3}\dot{h}_{\times}(t)\\
\\z(t)=l_{3}-\frac{1}{4[}(l_{1}^{2}-l_{2}^{2})\dot{h}_{+}(t)+2l_{1}l_{2}\dot{h}_{\times}(t).\end{array}\label{eq: Grishuk 0}\end{equation}

In absence of GWs the position of the mass is $(l_{1},l_{2},l_{3}).$
The effect of the GW is to drive the mass to have oscillations. Thus,
in general, from eqs. (\ref{eq: Grishuk 0}) all three components
of motion are present \cite{key-21,key-22}. Neglecting the terms
with $\dot{h}_{+}$ and $\dot{h}_{\times}$ in eqs. (\ref{eq: Grishuk 0}),
the {}``traditional'' equations for the mass motion are obtained
\cite{key-21,key-22,key-23,key-24,key-25}

\begin{equation}
\begin{array}{c}
x(t)=l_{1}+\frac{1}{2}[l_{1}h_{+}(t)-l_{2}h_{\times}(t)]\\
\\y(t)=l_{2}-\frac{1}{2}[l_{2}h_{+}(t)+l_{1}h_{\times}(t)]\\
\\z(t)=l_{3}.\end{array}\label{eq: traditional}\end{equation}

Clearly, this is the analogous of the electric component of motion
in electrodynamics \cite{key-21,key-22}, while equations

\begin{equation}
\begin{array}{c}
x(t)=l_{1}+\frac{1}{2}l_{1}l_{3}\dot{h}_{+}(t)+\frac{1}{2}l_{2}l_{3}\dot{h}_{\times}(t)\\
\\y(t)=l_{2}-\frac{1}{2}l_{2}l_{3}\dot{h}_{+}(t)+\frac{1}{2}l_{1}l_{3}\dot{h}_{\times}(t)\\
\\z(t)=l_{3}-\frac{1}{4[}(l_{1}^{2}-l_{2}^{2})\dot{h}_{+}(t)+2l_{1}l_{2}\dot{h}_{\times}(t),\end{array}\label{eq: news}\end{equation}

are the analogous of the magnetic component of motion. One could think
that the presence of these {}``magnetic'' components is a {}``frame
artefact'' due to the transformation (\ref{eq: trasf. coord.}),
but in Section 4 of \cite{key-21} eqs. (\ref{eq: Grishuk 0}) have
been directly obtained from the geodesic deviation equation too, thus
the magnetic components have a real physical significance. The fundamental
point of \cite{key-21,key-22} is that the {}``magnetic'' components
become important when the frequency of the wave increases but only
in the low-frequency regime. This can be understood directly from
eqs. (\ref{eq: Grishuk 0}). In fact, using eqs. (\ref{eq: onda generale})
and (\ref{eq: trasf. coord.}), eqs. (\ref{eq: Grishuk 0}) become

\begin{equation}
\begin{array}{c}
x(t)=l_{1}+\frac{1}{2}[l_{1}h_{+}(t)-l_{2}h_{\times}(t)]+\frac{1}{2}l_{1}l_{3}\omega h_{+}(t-\frac{\pi}{2})+\frac{1}{2}l_{2}l_{3}\omega h_{\times}(t-\frac{\pi}{2})\\
\\y(t)=l_{2}-\frac{1}{2}[l_{2}h_{+}(t)+l_{1}h_{\times}(t)]-\frac{1}{2}l_{2}l_{3}\omega h_{+}(t-\frac{\pi}{2})+\frac{1}{2}l_{1}l_{3}\omega h_{\times}(t-\frac{\pi}{2})\\
\\z(t)=l_{3}-\frac{1}{4[}(l_{1}^{2}-l_{2}^{2})\omega h_{+}(t-\frac{\pi}{2})+2l_{1}l_{2}\omega h_{\times}(t-\frac{\pi}{2}).\end{array}\label{eq: Grishuk 01}\end{equation}

Thus, the terms with $\dot{h}_{+}$ and $\dot{h}_{\times}$ in eqs.
(\ref{eq: Grishuk 0}) can be neglected only when the wavelength goes
to infinity, while, at high-frequencies, the expansion in terms of
$\omega l_{i}l_{j}$ corrections, with $i,j=1,2,3,$ breaks down \cite{key-21,key-22}.
Now, let us compute the total response functions of interferometers
for the {}``magnetic'' components. Equations (\ref{eq: Grishuk 0}),
that represent the coordinates of the mirror of the interferometer
in presence of a GW in the frame of the local observer, can be rewritten
for the pure magnetic component of the $+$ polarization as

\begin{equation}
\begin{array}{c}
x(t)=l_{1}+\frac{1}{2}l_{1}l_{3}\dot{h}_{+}(t)\\
\\y(t)=l_{2}-\frac{1}{2}l_{2}l_{3}\dot{h}_{+}(t)\\
\\z(t)=l_{3}-\frac{1}{4}(l_{1}^{2}-l_{2}^{2})\dot{h}_{+}(t),\end{array}\label{eq: Grishuk 1}\end{equation}

where $l_{1},l_{2}\textrm{ }and\textrm{ }\textrm{ }l_{3}$ are the
unperturbed coordinates of the mirror. To compute the response functions
for an arbitrary propagating direction of the GW, we recall that the
arms of the interferometer are in general in the $\overrightarrow{u}$
and $\overrightarrow{v}$ directions, while the $x,y,z$ frame is
adapted to the propagating GW (i.e. the observer is assumed located
in the position of the beam splitter) \cite{key-21,key-22}. Then,
a spatial rotation of the coordinate system has to be performed

\begin{equation}
\begin{array}{ccc}
u & = & -x\cos\theta\cos\phi+y\sin\phi+z\sin\theta\cos\phi\\
\\v & = & -x\cos\theta\sin\phi-y\cos\phi+z\sin\theta\sin\phi\\
\\w & = & x\sin\theta+z\cos\theta,\end{array}\label{eq: rotazione magn}\end{equation}

or, in terms of the $x,y,z$ frame:

\begin{equation}
\begin{array}{ccc}
x & = & -u\cos\theta\cos\phi-v\cos\theta\sin\phi+w\sin\theta\\
\\y & = & u\sin\phi-v\cos\phi\\
\\z & = & u\sin\theta\cos\phi+v\sin\theta\sin\phi+w\cos\theta.\end{array}\label{eq: rotazione 2 magn}\end{equation}

In this way the GW is propagating from an arbitrary direction $\overrightarrow{r}$
to the interferometer (see figure 2 in \cite{key-22} ). As the mirror
of eqs. (\ref{eq: Grishuk 1}) is situated in the \$u\$ direction,
using eqs. (\ref{eq: Grishuk 1}), (\ref{eq: rotazione magn}) and
(\ref{eq: rotazione 2 magn}) the $u$ coordinate of the mirror is
given by

\begin{equation}
u=L+\frac{1}{4}L^{2}A\dot{h}_{+}(t),\label{eq: du magn}\end{equation}

where

\begin{equation}
A\equiv\sin\theta\cos\phi(\cos^{2}\theta\cos^{2}\phi-\sin^{2}\phi)\label{eq: A}\end{equation}

and $L=\sqrt{l_{1}^{2}+l_{2}^{2}+l_{3}^{2}}$ is the length of the
interferometer arms.

The computation for the $v$ arm is similar to the one above. Using
eqs. (\ref{eq: Grishuk 1}), (\ref{eq: rotazione magn}) and (\ref{eq: rotazione 2 magn}),
the coordinate of the mirror in the $v$ arm is

\begin{equation}
v=L+\frac{1}{4}L^{2}B\dot{h}_{+}(t),\label{eq: dv magn}\end{equation}

where

\begin{equation}
B\equiv\sin\theta\sin\phi(\cos^{2}\theta\cos^{2}\phi-\sin^{2}\phi).\label{eq: B}\end{equation}

Equations (\ref{eq: du magn}) and (\ref{eq: dv magn}) represent
the distance of the two mirrors of the interferometer from the beam-splitter
in presence of the GW (note that only the contribution of the magnetic
component of the $+$ polarization of the GW is taken into account).
A {}``signal'' can also be defined in the time domain ($T=L$ in
our notation)

\begin{equation}
\frac{\delta T(t)}{T}\equiv\frac{u-v}{L}=\frac{1}{4}L(A-B)\dot{h}_{+}(t).\label{eq: signal piu}\end{equation}

The quantity (\ref{eq: signal piu}) can be computed in the frequency
domain using the Fourier transform of $h_{+}$, defined by

\begin{equation}
\tilde{h}_{+}(\omega)=\int_{-\infty}^{\infty}dth_{+}(t)\exp(i\omega t),\label{eq: trasformata di fourier magn}\end{equation}

obtaining

\[
\frac{\tilde{\delta}T(\omega)}{T}=H_{magn}^{+}(\omega)\tilde{h}_{+}(\omega),\]

where the function

\begin{equation}
\begin{array}{c}
H_{magn}^{+}(\omega)=-\frac{1}{8}i\omega L(A-B)=\\
\\=-\frac{1}{4}i\omega L\sin\theta[(\cos^{2}\theta+\sin2\phi\frac{1+\cos^{2}\theta}{2})](\cos\phi-\sin\phi)\end{array}\label{eq: risposta totale magn}\end{equation}

is the total response function of the interferometer for the magnetic
component of the $+$ polarization.

The analysis for the magnetic component of the $\times$ polarization
is similar. At the end one gets (see \cite{key-22} for details)

\begin{equation}
\begin{array}{c}
H_{magn}^{\times}(\omega)=-i\omega T(C-D)=\\
\\=-i\omega L\sin2\phi(\cos\phi+\sin\phi)\cos\theta.\end{array}\label{eq: risposta totale 2 per magn}\end{equation}
These response functions increase with increasing frequency. Thus,
one understands why the {}``magnetic'' components of GWs are important:
because of the increasing with increasing frequency, if one neglects
the {}``magnetic'' contributions, a portion of about the $15\%$
of the signal could be, in principle, lost in the high frequency portion
of the ground based GWs interferometers.

\subsection*{Conclusions}

After an introduction in which the importance of the gravitomagnetic
field in general relativity has been emphasized, with a resume of some astronomical and astrophysics applications,
the so-called {}``magnetic'' components of GWs, which have to be
taken into account in the context of the total response functions
of interferometers for GWs propagating from arbitrary directions,
have been discussed. We have shown that the magnetic contributions
turn out to be important for the total interferometric response functions
of both of the polarizations. In fact, if one neglects such contributions,
a portion of about the $15\%$ of the signal could be lost in the
high frequency portion of interferometers.

\end{document}